\documentclass[prd,aps,eqsecnum,floatfix,nofootinbib,preprint,tightenlines]{revtex4-2}
\usepackage{dcolumn}
\usepackage{amsmath}
\usepackage{latexsym}
\usepackage{graphicx}
\usepackage{bm}
\usepackage{colordvi}
\usepackage{xcolor}

\begin{document}
\title{The Ginsparg-Wilson relation and overlap fermions}
\author{Thomas DeGrand}
\affiliation{Department of Physics,
University of Colorado, Boulder, CO 80309, USA}
\email{thomas.degrand@colorado.edu}

\date{\today}

\begin{abstract}
I review the physics of lattice fermions obeying the Ginsparg-Wilson relation. I describe their relation
to domain wall fermions. I give a description of methodology for performing numerical simulations with overlap
fermions.
This is a chapter contributed to the on-line book ``Lattice QCD at 50 years,''
(LQCD@50), edited by
Tanmoy Bhattacharya,
Maarten Golterman,
Rajan Gupta,
Laurent Lellouch,
and
Steve Sharpe.
\end{abstract}
\maketitle

\section{Introduction and motivation}
The simplest lattice fermions have issues with chiral symmetry.
The situation is encoded in a  famous ``no-go'' theorem due to
Nielsen and Ninomiya \cite{Nielsen:1980rz,Nielsen:1981xu}, see also Karsten and Smit \cite{Karsten:1980wd}. The theorem makes two assumptions:
\begin{enumerate}
\item
There is a quadratic fermion action $\bar \psi(x)i H(x-y)\psi(y)$, where $H(p)$ is
continuous and well behaved. It should reduce to $\gamma_\mu p_\mu$ for
small $p_\mu$.
\item
There is a local conserved charge $Q$ defined as $Q=\sum_x j_0(x)$, which is quantized
(i.~e. $Q$ doesn't change across the Brillouin zone)
\end{enumerate}
The theorem is that, once these conditions hold,
$H(p)$ has an equal number of left handed and right handed fermions
for each eigenvalue of $Q$: this is called ``doubling.''
The choices we seem to have are to
work with fermion actions which are chiral but doubled (naive or staggered fermions)
or undoubled but with explicit order $a$  or $a^2$ violations of chiral symmetry (Wilson or clover fermions).

However there is a third way to evade the theorem, which basically amounts to making a redefinition 
of what is called ``chiral symmetry'' on the lattice. Fermion actions which do this are called 
``Ginsparg-Wilson'' actions (after their discoverers \cite{Ginsparg:1981bj}) or ``overlap'' actions (for their particular
implementation choice).
They are related to ``domain wall fermions,'' which live in five dimensions.
Domain wall fermions are described by Blum and Shamir in a separate review.
The connection is also described in Sec.~\ref{sec:overlapdwf} of this chapter.
Oddly enough, though, nearly all the technical developments of Ginsparg-Wilson fermions
and domain wall fermions proceeded in an essentially decoupled way.
This chapter attempts to give a stand-alone survey of lattice implementations of overlap fermions,
from a description of their formal background to a discussion of issues 
associated with their practical implementation in simulations.


\section{Formal properties of Ginsparg-Wilson fermions}

Overlap (or, more generally, Ginsparg-Wilson) fermions replace 
\cite {Luscher:1998pqa} the usual
definition of a chiral rotation
 $\delta \psi = i \epsilon \gamma_5 \psi$, $\delta \bar \psi = i \epsilon \bar
\psi \gamma_5$ by either
\begin{equation}
\delta \psi = i\epsilon \gamma_5\left(1-\frac{a}{2r_0}D\right) \psi; \qquad
\delta \bar \psi = i\epsilon  \bar \psi \left(1-\frac{a}{2r_0}D\right)\gamma_5
\end{equation}
or
\begin{equation}
\delta \psi = i\epsilon \gamma_5\left(1-\frac{a}{r_0}D\right) \psi; \qquad
\delta \bar \psi = i\epsilon\bar\psi\gamma_5.
\end{equation}
The operator $D$ should be local. The lattice spacing is $a$ and $r_0$ is a free parameter. The
naive $a\rightarrow 0$ limit of the new chiral rotation is just the
usual one, so this is just the ordinary chiral rotation modified by the
addition of an irrelevant operator.  In all situations I have seen, $D$ in
the rotation is taken  to be the (massless) Dirac operator used in the action, so that the
Lagrange density is ${\cal L}= \bar \psi D \psi$. Requiring $\delta
{\cal L}=0$ under either of the altered chiral rotations replaces the
usual anti-commutation relation of the Dirac operator and $\gamma_5$ by the new
constraint
\begin{equation}
 0= \gamma_5 D + D \gamma_5 - \frac{a}{r_0}D \gamma_5 D.
\label{eq:GW}
\end{equation}
This expression is called the ``Ginsparg-Wilson relation,'' named
after its discoverers \cite{Ginsparg:1981bj}. 

Ginsparg and Wilson discovered it as a natural consequence of performing 
a real-space renormalization group transformation on a fermion action.
Their Ginsparg-Wilson relation was $\{\gamma_5,D\} = a D\gamma_5 R D $ where $R$ is a
local operator. This amounts to writing the chiral rotation
as  $\delta \psi = i \epsilon \gamma_5 (1-aRD)\psi$. This
extension did not receive widespread use in practice.

The Ginsparg-Wilson relation can be rewritten as
\begin{equation}
\{\gamma_5, D^{-1}\} = \frac{a}{r_0} \gamma_5
\end{equation}
and, using $D^\dagger = \gamma_5 D \gamma_5$, a third version of the relation is
\begin{equation}
D + D^\dagger = \frac{a}{r_0}D^\dagger D.
\label{eq:GW2}
\end{equation}

Fermions obeying the Ginsparg-Wilson relation have a number of
remarkable properties. First, their eigenvalues are confined to a
circle of radius $r_0/a$ in the complex plane, centered at the point
$(r_0/a,0)$.  To see this, write the eigenvalue equation
as $D \phi = \lambda \phi$ with $\lambda = x+iy$.  Eq.~\ref{eq:GW2}
says that $2x = (a/r_0)(x^2 + y^2)$ or
$(x-r_0/a)^2 + y^2 = (r_0/a)^2$.  Second, if an eigenmode of $D$ is
chiral, $\gamma_5 \phi = \pm \phi$, then $\lambda$ is real and equal
to either 0 or $2r_0/a$,
\begin{equation}
(\gamma_5 D + D \gamma_5 - \frac{a}{r_0}D\gamma_5 D)\phi =
(\pm 2 \lambda  \mp \frac{a}{r_0}\lambda^2)\phi ,
\end{equation}
so $ \lambda[2 -(a/r_0)\lambda]=0$.

The Hermitian Dirac operator $H = \gamma_5 D$ is another useful object.
In terms of $H$, Eq.~(\ref{eq:GW2}) becomes
\begin{equation}
\gamma_5 H + H \gamma_5 = \frac{a}{r_0}H^2.
\label{eq:GW3}
\end{equation}
If an eigenstate $\mathinner{|{\psi}\rangle}$ of $H$ is also an eigenstate of $\gamma_5$,  then from
Eqn.~\ref{eq:GW3}, its $H$ eigenvalue must be one of the possibilities $(0,\pm 2r_0/a)$.
Then it is an eigenstate of all four operators $H$, $H^2$, $D$, and
$\gamma_5$.  Otherwise, $\mathinner{|{\psi}\rangle}$ and $\gamma_5 \mathinner{|{\psi}\rangle}$ form a
two-dimensional subspace that is invariant under the action of $H$ and
$\gamma_5$ (and consequently $D$), and on which $H$ has eigenvalues
$\pm\epsilon$.

This is easiest to see by
beginning with an eigenvector of $H$, which obeys
$H|\psi\rangle = \epsilon |\psi\rangle$ for $\epsilon \ne 0$ or $\pm 2r_0/a$,
and constructing the orthogonal state
\begin{equation}
|\phi\rangle = N[ \gamma_5|\psi\rangle - |\psi\rangle\langle\psi|\gamma_5|\psi\rangle],
\end{equation}
where $N$ is a normalization factor, $1/N^2= 1 - |\langle\psi|\gamma_5|\psi\rangle|^2$.
(The construction assumes that $\mathinner{|{\psi}\rangle}$ is not
 an eigenstate of $\gamma_5$.)
Computing $H|\phi\rangle$ and using the Ginsparg-Wilson relation, Eq.~(\ref{eq:GW3}), we get
\begin{equation}
H|\phi\rangle = -\epsilon|\phi\rangle + |\psi\rangle N\epsilon
  \left(\epsilon\frac{a}{r_0} - 2\langle\psi|\gamma_5|\psi\rangle \right).
\end{equation}
So we see that the subspace is invariant under $H$.
From this we learn that
$|\phi\rangle$ is an eigenstate of $H$ with eigenvalue $-\epsilon$ and (because $H$ is Hermitian)
$\langle\psi|\gamma_5|\psi\rangle=\epsilon/(2r_0/a)$.
Finally, we see that $H^2$ is just $\epsilon^2$ on this subspace and it
commutes with $\gamma_5$, so we
can diagonalize $H^2$ and $\gamma_5$ simultaneously.

In a basis where $\gamma_5 = {\rm diag}(+1,-1)$, we can write
\begin{equation}
\psi= \left(\begin{array}{c} \alpha \\ \beta \end{array} \right) = 
\frac{1}{\sqrt{2}} \left( \begin{array}{c} \sqrt{1+\frac{\epsilon a}{2r_0}} \\ \sqrt{1-\frac{\epsilon a}{2r_0}} \end{array} \right) .
\end{equation}
This allow us to write down $H$ in the simultaneous
eigenbasis of $H^2$ and $\gamma_5$. It is
a direct sum of two by two blocks, each of which is
\begin{equation}
H = \frac{1}{2r_0/a} \left(
\begin{array}{cc} \epsilon^2 & \epsilon\sqrt{(2r_0/a)^2-\epsilon^2} \\
        \epsilon\sqrt{(2r_0/a)^2-\epsilon^2} & -\epsilon^2 \end{array} \right).
        \label{eq:heig}
\end{equation}
In addition $H$ may
have simple diagonal terms for $\epsilon = 0$ or $2 r_0/a$.

This seems formal but it is quite useful when one wants to find eigenfunctions of
$D$ or $H$: the spectrum of $H^2$ is bounded from below, so numerical methods which find
the lowest eigenvalue eigenmodes of a matrix can be used, and one can construct as much of the
spectrum as one wants working in a single chirality basis. Typically, this
is the sector which contains all the zero modes.

The operator $D = \gamma_5 H$ reduces, of course, to block-diagonal form
on the same
two- or one-dimensional eigenspaces of $H^2$.
Where the eigenvalues of $H$ are $\pm
\epsilon$, the eigenvalues of $D$ come in complex conjugate pairs,
\begin{equation}
\lambda = \frac{\epsilon^2 \pm i \epsilon\sqrt{(2r_0/a)^2-\epsilon^2}}{2r_0/a},
\end{equation}
or they are real: $\epsilon = 0$ corresponds to $\lambda = 0$ and
$\epsilon = \pm 2r_0/a$ corresponds to $\lambda = 2r_0/a$.
Where eigenmodes of $H$ with nonzero eigenvalue $\pm \epsilon$ have
matrix elements
 $\langle \phi | \gamma_5 | \phi \rangle = \pm\epsilon/(2r_0/a)$,
 from Eq.~\ref{eq:GW} it is easy to show that nonzero
eigenvalue eigenmodes of $D$ have
 $\langle \psi | \gamma_5|\psi\rangle = 0$.

Since the Hilbert space of Dirac fields is complete and finite
dimensional it must be that $\mbox{Tr} \; \gamma_5 = 0$.  Taking the trace on
the eigenbasis of $D$, we find that the only nonzero contributions
come from the eigenspaces $0$ and $2r_0/a$.  In the eigenspace $0$,
$\mbox{Tr} \; \gamma_5$ counts the difference $n_+ - n_-$ between zero
eigenmodes of positive and negative chirality.  In the eigenspace
$2r_0/a$ it counts the corresponding difference $n_+'-n_-'$.  For the
net trace to vanish, we must have $(n_+-n_-)+(n_+'-n_-')= 0$.
Similar reasoning
shows that $\mbox{Tr} \; \gamma_5 D = (2r_0/a )(n_+' - n_-')$.  This leads us to the identity
\begin{equation}
  Q = \frac{a}{2r_0}\mbox{Tr} \; \gamma_5 D = - \mbox{Tr} \; \gamma_5\left( 1-\frac{a}{2r_0}D\right) = n_- - n_+  .
\end{equation}
This is the first part of the index theorem, which equates the topological
charge (winding number) to $Q$, the difference between the numbers of zero eigenmodes
of positive and negative chirality, gauge configuration by
configuration.  To complete the theorem it is necessary to relate $Q$ to the
topological charge.  For a smooth background gauge configuration,
showing that $\mbox{Tr} \; \gamma_5 D $ is proportional to the integral over
$\epsilon^{\alpha\beta\mu\nu}F^a_{\alpha\beta}F^a_{\mu\nu}$ involves
manipulations similar to what is done in the continuum calculation
(and for that, see the text \cite{Peskin:1995ev}).

It is convenient  to define the
massive overlap operator for fermion mass $m$ as
\begin{equation}
D(m) = \left(1-\frac{m}{2r_0/a}\right)D + m.
\label{eq:dm}
\end{equation}
With this definition, (naturally $H(m)=\gamma_5 D(m)$),
\begin{equation}
D(m)^\dagger D(m) = H(m)^2= \left[ 1-\frac{m^2}{4(r_0/a)^2}\right] H^2 +m^2 .
\label{eq:hm2}
\end{equation}

The ``shifted propagator'' $\hat D(m)^{-1}$ is often used in
calculations:
\begin{equation}
\left(1-\frac{m}{2r_0/a}\right)\hat D(m)^{-1} = D(m)^{-1} - \frac{1}{2(r_0/a)}.
\label{eq:shift}
\end{equation}
It obeys
\begin{equation}
\{ \gamma_5, \hat D(m)^{-1} \} = 2m \hat D(m)^{-1} \gamma_5 \hat D(m)^{-1}.
\label{eq:useful}
\end{equation}
 $\hat D$, the inverse
shifted propagator, is generally not local, but its anti-commutator is
quite familiar: $\{ \gamma_5, \hat D \} = 2m \gamma_5$.
One could not ask for anything more chiral than this.

For the domain wall story about $\hat D(m)^{-1}$, see the chapter by Blum and Shamir, Sec. 4.5.

Eq.~\ref{eq:useful} immediately gives us $m_{PS}^2\propto m_q$: 
multiply both sides on the left by  $\gamma_5$, trace and sum over sites, 
\begin{equation}
2{\rm Tr} \hat D^{-1}= 2m   {\rm Tr}  \gamma_5 \hat D(m)^{-1} \gamma_5  \hat D(m)^{-1}.
\label{eq:pregmor}
\end{equation}
The right hand side has the topology of a flavor nonsinglet pseudoscalar correlator. Eq.~\ref{eq:pregmor} can be
written as
\begin{equation}
 \frac{1}{V}\sum_x \left\langle{S(x)}\right\rangle = \frac{m_q}{V} \sum_{x,y} \left\langle{P(x) P(y)}\right\rangle
\label{eq:gmor}
\end{equation}
with the scalar and pseudoscalar currents defined (to leading order in $m$) as
\begin{equation}
S= \bar q(x,t) (1-\frac{a}{2r_0} D) q(x,t); \quad P= \bar q(x,t)  (1-\frac{a}{2r_0} D)   \gamma_5  q(x,t).
\end{equation}
When chiral symmetry is broken, the left hand side of Eq.~\ref{eq:gmor} is the chiral condensate $\Sigma$.
Saturating the correlator on the right hand side  of Eq.~\ref{eq:gmor} with the pseudoscalar state gives us 
\begin{equation}
C(t)= \sum_{\vec x} \left\langle{P(\vec x ,t) P(0,0)}\right\rangle = |\left\langle{0|P|PS}\right\rangle|^2 \frac{\cosh (m_{PS}(N_t/2-t))}{2m_{PS}\sinh (m_{PS}N_t/2)} .
\end{equation}
and its integral over the simulation volume is
\begin{equation}
\int_0^{N_t} dt C(t)= \frac{ |\left\langle{0|P|PS}\right\rangle|^2 }{m_{PS}^2} .
\label{eq:condsub}
\end{equation}
When $\Sigma$ is nonzero this gives us
\begin{equation}
 \frac{ |\left\langle{0|P|PS}\right\rangle|^2 }{m_{PS}^2} =   \frac{\Sigma}{m_q} .
\end{equation}
which is $m_{PS}^2 \propto m_q$.

Chiral Ward identities come from writing Eq.~\ref{eq:useful} as
\begin{equation}
\hat D(m)^{-1} + \gamma_5 \hat D(m)^{-1} \gamma_5 = 2m \gamma_5 \hat D(m)^{-1} \gamma_5 \hat D(m)^{-1}
\end{equation}
so
\begin{eqnarray}
{\rm Tr}  \hat D(m)^{-1} \Gamma   \hat D(m)^{-1} \Gamma +
{\rm Tr}  \hat D(m)^{-1} \Gamma  (\gamma_5  \hat D(m)^{-1}  \gamma_5) \Gamma  \nonumber \\
=2m {\rm Tr}  \hat D(m)^{-1} \Gamma  (\gamma_5  \hat D(m)^{-1}  \gamma_5  \hat D(m)^{-1} \gamma_5) \Gamma \nonumber\\
\label{eq:chward}
\end{eqnarray}
for any gamma matrix $\Gamma$. For $\Gamma=\gamma_\mu$ this is the vector - axial vector - pseudoscalar Ward identity
\begin{equation}
\left\langle{V_\mu V_\mu} \right\rangle - \left\langle{A_\mu A_\mu} \right\rangle = 2m \left\langle{A_\mu PV_\mu} \right\rangle.
\label{eq:vecward}
\end{equation}
for
\begin{equation}
V_\mu= \bar q(x,t) (1-\frac{a}{2r_0} D) \gamma_\mu  q(x,t); \quad A_\mu=  \bar q(x,t)  (1-\frac{a}{2r_0} D)  \gamma_\mu \gamma_5  q(x,t).
\end{equation}
None of the currents we have defined are Noether currents, so there will be matching factors connecting them to
their continuum analogs. The continuum Ward identities are identical to the ones we have written.
Eq.~\ref{eq:vecward},  plus the corresponding one with $\Gamma=1$, tell us 
that lattice to continuum matching factors for the vector
and axial currents
are equal, as are those for the scalar and pseudoscalar
currents, $Z_V=Z_A$, $Z_P=Z_S$.

Notice that the extra terms in the currents and identities
 compared to the continuum relations involve the product $D^{-1} D$, which is a contact term in the correlation functions.

The singlet analog of Eq.~\ref{eq:pregmor} includes the topological charge and gives the 
Di Vecchia and Veneziano
 \cite{DiVecchia:1980yfw},  Leutwyler and Smilga \cite{Leutwyler:1992yt}, and Crewther \cite{Crewther:1977ce} relation,
 that the
 topological susceptibility vanishes linearly with the fermion mass. For details, see (for example)
 the early review by  Niedermeyer \cite{Niedermayer:1998bi}.

This is all very magical: we have a lattice fermion whose chiral properties
are equal (up to contact terms) to those of continuum fermions, without doubling.
To quote from a letter to the author from Peter Hasenfratz: ``It is like
opening Pandora's box.''

 The final issue is, what is an explicit formula for
a Dirac operator which obeys the Ginsparg-Wilson relation? 
One last definition gives a clue:
Because the eigenvalues of $D$ lie on a shifted circle, one can write $D$ as
\begin{equation}
D = \frac{r_0}{a}[1+V]
\end{equation}
where $V$ is a unitary operator. We can choose $V$ to be a
function of some ordinary, nonchiral, undoubled massless lattice
Dirac operator $d$ (referred to hereafter as the ``kernel'' Dirac operator)
and a mass term equal to $-r_0/a$.  The
universally-used  choice of function is the overlap action of Refs.~\cite{Neuberger:1997fp,Neuberger:1998my}
\begin{equation}
 D= \frac{r_0}{a}\left[1 + \frac{d-r_0/a}{\sqrt{(d-r_0/a)^\dagger (d-r_0/a)}}\right]
\label{eq:overlap}
\end{equation}
or, introducing the Hermitian operator $h(m)=\gamma_5(d+m)$,
\begin{equation}
D= \frac{r_0}{a}[1 +  \gamma_5 \epsilon[h(-r_0/a)]]
\label{eq:msf}
\end{equation}
where $\epsilon(h)$ is the matrix step function, $\epsilon(h) =h/\sqrt{h^2}$. 
The choice of coefficients is made for convenience so that the small eigenvalue limit the overlap operator has the
same normalization as the kernel operator
 $D \simeq d \simeq i \partial {\!\!\!\!/\,}$.

All magical things have a high price. Here, the problem is that $V$
is not restricted to extend over a finite number of lattice sites.
Said another way, it is not ultralocal.
This means that the matrix representation of $D$ is not sparse.

\begin{figure}
\begin{center}
\includegraphics[width=0.8\textwidth,clip]{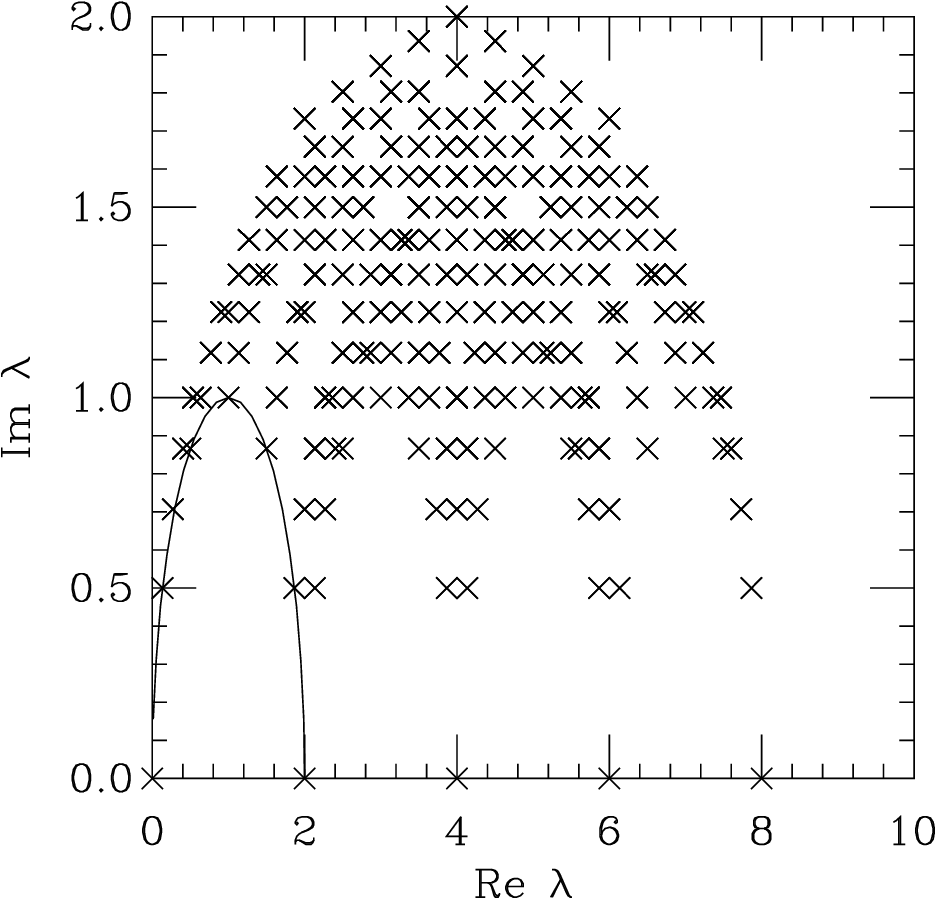}
\end{center}
\caption{
A comparison of the spectrum of eigenvalues of the ordinary Wilson fermion action (crosses)
on a finite lattice, with the corresponding spectrum of an $r_0/a=1$ overlap action (the line).
}
\label{fig:circlew}
\end{figure}

There was a theoretical issue in the literature: was $D$ local?
``Locality'' means that the size of
$D(|x-y|)$ dies exponentially with  the separation $|x-y|$,
 $D\propto \exp(-C|x-y|)$. Here $|x-y|$ is a distance measured in lattice units;
the physical distance is $r=a|x-y|$ and $D\propto \exp(-C r/a)$ vanishes when $r\gg a$.
The answer to this question depended on the choice of the kernel.
A proof of locality when the kernel $d$ was the Wilson action with thin links,
but in the limit of very smooth gauge fields, was presented in Ref.~\cite{Hernandez:1998et}.
Most of the literature involved issues related to using the Wilson action as a kernel.
I never did that, so for me the only issue of principle is illustrated in Fig.~\ref{fig:circlew}.
The overlap maps $d$ into a circle. All of the  eigenmodes of $d$  with real eigenvalues closer to the origin  than
the value of $r_0/a$ will be pushed to the origin and act as separate flavors. We don't want doubling,
so there had better only be one set of these modes. In the figure, it had better be that $r_0/a<1$.
This was never an issue in practice to maintain. Also, in practice, $C$ was never known from first principles,
so  everybody who ever simulated overlap fermions looked at a plot of something like
$D(x-y)\delta(y)$ as a function of $|x-y|$ and tuned $d$ and $r_0$ to minimize the range of $D$.

This takes us (after a brief aside) to Sec.~\ref{sec:ovmeth}, which is a discussion of
practical considerations: how to construct the overlap $D$ (and related quantities) as efficiently as possible.

\section{The overlap action beginning in five dimensions\label{sec:overlapdwf}}

We pause to return to the discussion of chiral fermions in five dimensions.
Let us re-state some conventions: the coordinate in the fifth dimension is $s$,
and the five-dimensional Dirac operator $D_5$ is written in 
terms of a four dimensional Dirac operator $D_4$ and a mass function $M(s)$
as
\begin{equation}
D_5=D_4 + \gamma_5 \partial_5 -M(s).
\end{equation}

Since the situation does not depend on a particular form for $M$, let us
replace it by a wall with Dirichlet boundary conditions at $s=0$.  We
restrict ourselves to $s>0$ and take $M(s)=M$ in that domain. For this
simplified case the left-handed zero mode is pinned to the $s=0$
surface and its wave function falls away like $\exp(-Ms)$. Now let us
find the four dimensional effective field theory of the massless
mode. We can do this by finding the propagator $G(x,s;y,s_0)$
between two sites in the five-dimensional manifold and restrict our
attention to the case $s=s_0=0$ (since that is where the massless
mode is confined).

The Green's function obeys
\begin{equation}
D_5 G(x,s;y,s_0)=\delta^4(x-y)\delta(s-s_0)
\end{equation}
with the boundary conditions that $G(x,s;y,s_0)$ should vanish as $|s-s_0|$ goes to infinity and
\begin{equation}
P_+G(x,s;y;s_0)=0
\label{eq:chbc}
\end{equation}
at $s=0$.  Here I introduce the chiral projectors
\begin{equation}
P_\pm=\frac{1}{2}(1\pm \gamma_5).
\end{equation}
A few more definitions are useful before we proceed. First, set
\begin{equation}
Q=\gamma_5(M-D_4)
\end{equation}
so
\begin{equation}
D_5=\gamma_5( \partial_5 -Q).
\end{equation}
To make contact with the notation of previous sections, $M=r_0/a$, $D_4=d$ 
is the kernel operator and $Q=-h(-r_0/a)$ is the Hermitian Dirac operator.
Eigenmodes of $Q$,
\begin{equation}
Q\mathinner{|{n}\rangle} =q_n\mathinner{|{n}\rangle},
\end{equation}
will be used to construct the solution.
Positive and negative eigenvalue projectors of $Q$, 
\begin{eqnarray}
\hat P_\pm &=& \frac{1}{2}(1\pm \frac{Q}{\sqrt{Q^2}}) \nonumber\\
& \equiv& \frac{1}{2}(1\pm\epsilon(Q))  ,  \nonumber\\
\end{eqnarray}
will  play a  r\^ole in what follows.

Our goal is to show that
\begin{equation}
G(x,s=0;y,s_0=0) = 2MP_- S(x,y)P_+,
\label{eq:gwM}
\end{equation}
where $S(x,y)=D_{ov}(x,y)^{-1}$ if $x \ne y$, with
\begin{equation}D_{ov}=M[1-\gamma_5\epsilon(Q) ].
\end{equation}
That is, $D_{ov}$ is the overlap Dirac operator.
The derivation of this result was first given by L\"uscher \cite{Luscher:2000hn}.

To begin,  write the four dimensional delta function in terms 
of the eigenmodes of $Q$
\begin{equation}\delta^4(x-y)=\sum_n \left\langle{x|n}\right\rangle \left\langle{n|y}\right\rangle
\label{eq:g2}
\end{equation}
and likewise write the Green's function as
\begin{equation}G(s,x;s_0,y) = \sum_n    \left\langle{x|n}\right\rangle    {\cal G}_n(s,s_0) \left\langle{n|y}\right\rangle .
\label{eq:g1}
\end{equation}
Then
${\cal G}_n(s,s_0)$ obeys the differential equation
\begin{equation}
\gamma_5(\partial_5-q_n){\cal G}_n(s,s_0) = \delta(s-s_0).
\label{eq:g4}
\end{equation}
Leaving aside the boundary conditions for the moment, the solution to this simple problem is
\begin{eqnarray}
{\cal G}_n(s,s_0) &=& [Ae^{q_n(s-s_0)} ]\gamma_5 \qquad  \rm{if} \ s<s_0 \nonumber\\
&=& [Be^{q_n(s-s_0)} ]\gamma_5 \qquad  \rm{if} \ s>s_0 \nonumber\\
\end{eqnarray}
where $B-A=1$ to handle the discontinuity due to the delta function.
At this point we can write the Green's function as
\begin{equation}
G(s,x;s_0,y) = \sum_n    \left\langle{x|n}\right\rangle  e^{q_ns}[ \theta(s-s_0) + C]   e^{-q_n s_0}    \gamma_5    \left\langle{n|y}\right\rangle       .
\label{eq:g3}
\end{equation}
(I have split the $\left\langle{x|n}\right\rangle  $ and $ \left\langle{n|y}\right\rangle $ apart because at the end I will want to write the answer directly in terms of
operators depending on $Q$.)

Now for the boundary conditions. We want the solution to fall to zero as $|s-s_0|$ goes to infinity. Make the replacement
\begin{equation}
\theta(s-s_0)= \theta(s-s_0) (\hat P_+ + \hat P_-)
\end{equation}
in Eq.~\ref{eq:g3}.
The $\hat P_-$ part of this expression gives a contribution to the Green's function which
 falls to zero at big $|s-s_0|$. To handle the positive eigenmodes, write
\begin{equation}
\theta(s-s_0)\hat P_+ = -(1-\theta(s-s_0))\hat P_+  + \hat P_+ .
\end{equation}
Now we have
\begin{eqnarray}
G(s,x;s_0,y) &=& \sum_n    \left\langle{x|n}\right\rangle  e^{q_n s}[ \theta(s-s_0)\hat P_-    \nonumber\\
 & &   -(1-\theta(s-s_0) )\hat P_+ + (\hat P_+ +C)  ]   e^{-q_n s_0}   \gamma_5     \left\langle{n|y} \right\rangle   . \nonumber\\
\label{eq:g11}
\end{eqnarray}
or
\begin{equation}
G(s,x;s_0,y) = \sum_n    \left\langle{x|n}\right\rangle  e^{q_n s}[ \theta(s-s_0)\hat P_-   -\theta(s_0-s) \hat P_+ +  C'  ]   e^{-q_n s_0}   \gamma_5   \left\langle{n|y}\right\rangle  .
\label{eq:g5}
\end{equation}
The first two terms satisfy the desired boundary condition at large $|s-s_0|$.
To make sure that the $C'$ term also obeys the boundary condition, we need $C' = \hat P_- O \hat P_+$
for some $O$.  We have arrived at
\begin{eqnarray}
G(s,x;s_0,y) &=& \sum_n    \left\langle{x|n}\right\rangle  e^{q_n s}[ \theta(s-s_0)\hat P_-    \nonumber\\
& &   -\theta(s_0-s) \hat P_+ +   \hat P_- O \hat P_+ ]   e^{-q_n s_0}    \gamma_5    \left\langle{n|y}\right\rangle   . \nonumber\\
\label{eq:g6}
\end{eqnarray}
Now we take the limit $s\rightarrow 0$ first, then $s_0\rightarrow 0$. In everything we have done so far, the $\hat P_\pm$'s were functions of the $q_n$'s, but we can
now interpret them as functions of the $Q$'s (which are themselves functions of the four dimensional coordinates) and strip off the explicit
dependence on the eigenmodes of $Q$.
Suppressing the labels $s$ and $s_0$, our Green's function is
\begin{equation}
G(x,y) = [ - \hat P_+ +   \hat P_- O \hat P_+  ]    \gamma_5     .
\label{eq:g9}
\end{equation}
On to the last boundary condition, Eq.~\ref{eq:chbc}. This is
\begin{equation}
(-P_+ + P_+ \hat P_- O) \hat P_+ = 0
\end{equation}
We use
\begin{equation}
P_+ \hat P_-  = \frac{1}{2}(1+\gamma_5)\frac{1}{2}(1-\epsilon) = \frac{1}{2}(1+\gamma_5)\frac{1}{2}(1- \gamma_5 \epsilon)  = P_+\frac{D_{ov}}{2M}
\end{equation}
which says that 
\begin{equation}
O= \frac{2M}{D_{ov}}.
\end{equation}
The Green's function is
\begin{equation}
G(x,y) = [  - \hat P_+ +   \hat P_-  \frac{2M}{D_{ov}}     \hat P_+  ]    \gamma_5     .
\label{eq:g7}
\end{equation}
Finally, the improbable identities (for the student to check!) 
  $\epsilon(Q)(M/D_{ov}) = \gamma_5(M/D_{ov} -1)$, 
$(M/D_{ov})\epsilon(Q) \gamma_5= \epsilon \gamma_5 + M/D_{ov}$,
and
$\epsilon(Q)(M/D_{ov})\epsilon(Q)\gamma_5=\gamma_5 (M/D_{ov})$
  give
\begin{equation}
G(x,y) =   P_-  + P_-  \frac{2M}{D_{ov}}  P_+      .
\label{eq:g8}
\end{equation}
which is what we wanted to show.
The $P_-$ term is a contact term, and we can neglect it.

\section{Numerical methods for overlap fermions \label{sec:ovmeth}}

Let's write down the overlap operator one more time and look at it from the point
of view of a programmer who wants to write efficient code to evaluate its
action on a trial vector:
\begin{equation}
  D= r_0[1+\gamma_5 \epsilon(h(-r_0))] .
\end{equation}
Again, $h(-r_0)=\gamma_5(d-r_0)$ is a kernel Hermitian Dirac operator and
$\epsilon(x)=x/\sqrt{x^2}$ is the matrix step function. Evaluating $\epsilon(h)$ is the real
bottleneck to using the overlap operator in a simulation.

One can approximate $\epsilon(h)$ either as a polynomial or as
a rational function  in $h$.    A polynomial approximation is a power series
\begin{equation}
  \epsilon(h) = h s_N(h^2) = h \sum_{n=0}^N c_n h^{2n},
\end{equation}
where $s_N(x)$ is an approximation to $1/\sqrt{x}$.  This can be (and has been)
implemented as a series expansion in Chebychev polynomials $T_n(x)$,
\begin{equation}
s_N(x) = \sum_n c_n T_n(x).
\end{equation}
Diagonal rational functions are more efficient.  In this case we write
\begin{eqnarray}
  \epsilon_N(h) &=& h \frac{\sum a_n h^{2n}}{\sum b_n h^{2n}} \nonumber \\
 & = & h\left[c_0 + \sum_{j=1}^N\frac{c_j}{h^2+d_j} \right],
\label{eq:epsgen}
\end{eqnarray}
where the $c$'s and $d$'s are related to the $a$'s and $b$'s. The second expression in Eq.~\ref{eq:epsgen}
is the one which is always used, since the
 sum of inverses can be computed using a multishift conjugate
gradient algorithm.

For real $x$,  any approximate $\epsilon_N(x)$ will have a range $0 <
x_{\rm min} < x < x_{\rm max}$ where it works reasonably well, but
outside that range, it will fail.  These end points can be adjusted by a
multiplicative change of scale, but the lower end point is never zero.
In order that $\epsilon_N(x)$ is to be a good approximation to the
$\epsilon(h)$ we need for the overlap, it must be  that the good interval contains all the
eigenvalues of $h$ from the smallest (absolute value) $|\lambda_{min}|$ to largest
$|\lambda_{\rm max}|$.  The largest eigenvalue is bounded by the lattice
cutoff, so in  typical simulations, its value is very stable. As usual, the smallest
eigenvalue is the problematic one.  Indeed, in a dynamical simulation,
$\lambda_{min}$ will cross the origin when the topology of the gauge
configuration changes, so one will always encounter arbitrarily small
values of $\lambda$.

The cure to this problem is to isolate the small eigenvalue eigenmodes and
include their contributions exactly.  Find them and sort the eigensolutions
$h(-r_0)|j\rangle = \lambda_j|j\rangle$ in order from the  smallest
$|\lambda_0|$ to largest  $|\lambda_{\rm max}|$.  Then apply $\epsilon(h)$ to a vector
$|\psi\rangle$ as follows:
\begin{equation}
 \epsilon(h)|\psi\rangle = \epsilon_N(h)\left(|\psi\rangle
   - \sum_{j=1}^J|j\rangle\langle j|\psi\rangle\right)
   + \sum_{j=1}^J|j\rangle \epsilon(\lambda_j)\langle j|\psi\rangle .
\label{eq:lowh}
\end{equation}
Now $\epsilon_N(h)$ has to be a good approximation only for the
range $|\lambda_{J+1}|$ to $|\lambda_{\rm max}|$. 

Clearly, some tuning is involved
to find an optimum choice of parameters. Eigenfunctions and eigenvalues of $h$ have to be found;
this is done by finding a set of low-lying spectrum of $h^2$ (which is bounded from below) and then
diagonalizing $h$ in the
eigenbasis. This is expensive and the cost grows as more and more eigenmodes are needed.
 Then one must find a good combination of $c_j$, $d_j$ and $N$ in Eq.~\ref{eq:epsgen}
and $J$ in Eq.~\ref{eq:lowh} to insure that the step function is being evaluated to some desired accuracy
and at minimum cost.

This can be tested in  simulation in several ways.
One can ask  whether $D$ or $H$ have chiral zero modes
(zero eigenvalue at the level of desired precision) on sample gauge 
configurations, such as ones with an isolated instanton.
One can ask the same question, but on gauge fields in the ensemble one hopes to use for physics.
Generally, this is a more stringent test, since the rougher the gauge configuration, the harder
it is for an approximation to the overlap to capture its properties.
Finally, one can check that one's results encode chiral symmetry, either 
by checking that the squared pion mass vanishes at zero bare quark mass, or by observing
the absence of additive mass renormalization
 via a measurement of  the Axial Ward Identity 
fermion mass involving the axial current
 and the pseudoscalar current,
\begin{equation}
\partial_t \sum_{\vec x} \left\langle{A_0(\vec x,t)O(0,0)}\right\rangle = 2am_q^{AWI} \sum_{\vec x} \left\langle{ P^a(\vec x,t)O(0,0)}\right\rangle.
\label{eq:AWI}
\end{equation}
This  choice is easier to implement and interpret.

Two examples of $\epsilon_N$'s are the polar expression and the
Zolotarev approximation.  The polar expression [referring to
Eq.~(\ref{eq:epsgen})] has $c_0=0$ and
\begin{equation}
 c_k  =  \frac{1}{N\cos^2\left[\pi(k-\frac{1}{2})/2N \right]} ;\qquad
 d_k  =  \tan^2\left[\pi(k-{\frac{1}{2}})/2N\right]
\end{equation}
It is actually the power series expansion of the approximate step
function used by the ordinary finite-fifth-dimension domain wall fermion.
In my experience, it is not robust enough to use in most situations.

The Zolotarev  formula represents the state of the art.
Referring readers to the mathematical literature (see ~\cite{Akhiezer1,Akhiezer2})   for its derivation,
I simply present the recipe for its implementation. The coefficients are defined in terms of the elliptic
integrals ${\rm cel}$, ${\rm cn}$, ${\rm sn}$, and ${\rm dn}$,
which I will list due to the multitude of conventions (I am following those of Ref.~\cite{Press:1988xx}):
The  approximate step function is
\begin{equation}
  \epsilon_N(x) =  x\sum_{\ell=1}^N\frac{b_\ell}{x^2+c_{2\ell-1}} .
\end{equation}
For $x \in (x_{\rm min},x_{\rm max})$ define $\kappa= x_{\rm min}/x_{\rm max}$;
$\kappa_p= {\rm cel}(\kappa,1,1,1)$; and $u=\ell\kappa_p/(2N)$.  Then the
partial fraction coefficients are
\begin{eqnarray}
  c_\ell &=& \left[\frac{{\rm sn}(u,\kappa^2)}{{\rm cn}(u,\kappa^2)}\right]^2
      x_{\min}^2 \nonumber \\
  b_\ell &=& \frac{\prod_{i=1}^{N-1}(c_{2i}-c_{2\ell-1})}
     { \prod_{i=1; i\ne \ell}^{N}(c_{2i-1}-c_{2\ell-1})} .
\end{eqnarray}
The elliptic integrals are
\begin{eqnarray}
{\rm cel}(k,p,a,b) &=&\int_0^\infty \frac{a+bx^2}{(1+px^2)\sqrt{(1+x^2)(1+k^2x^2)}} dx
\nonumber \\
y &=&{\rm sn}(u,k)\nonumber\\
 \qquad u &=& \int_0^{{\rm sn}}\frac{dy}{\sqrt{(1-y^2)(1-k^2y^2)}}
\end{eqnarray}
with ${\rm sn}^2 + {\rm cn}^2=1$, $k^2 {\rm sn}^2+{\rm dn}^2=1$.

\begin{figure}
\begin{center}
\includegraphics[width=0.8\textwidth,clip]{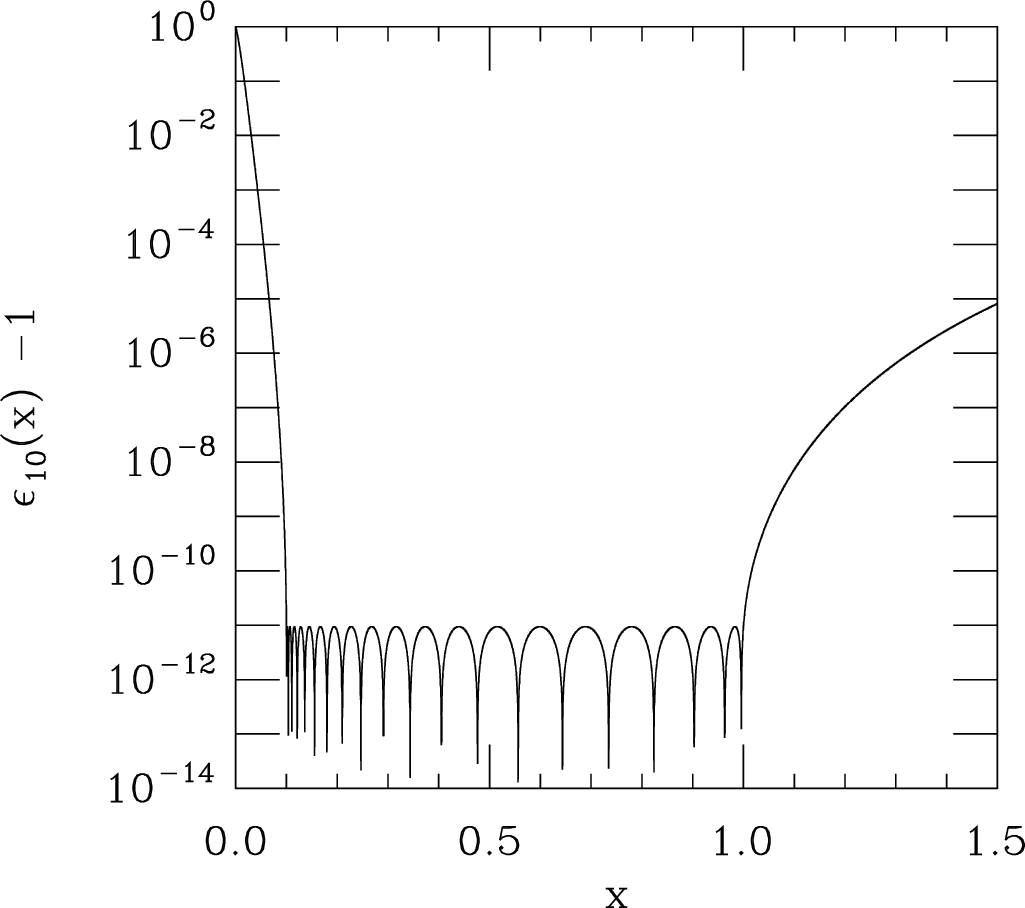}
\end{center}
\caption{
An example of a tenth-order Zolotarev approximation to the step function, where the desired
range is for $0.1<x<1$. We plot $\epsilon_{10}(x)-1$ to display the error in the
approximation.
}
\label{fig:zol}
\end{figure}

 The extrema of $\epsilon_N$ are located at $x_1=x_{\min}$ and
$x_2=x_{\rm min}/{\rm dn}[\kappa_p/(2N)]$.
The resulting $\epsilon_N(x)$ needs a small correction to remove an
asymmetric deviation from $\epsilon(x) = 1$. There does not seem to be
 a preferred minimization criterion (absolute deviation, least
squares, etc.) in the literature.   Replacing $b_\ell$ by
$2b_\ell/|\epsilon_N(x_1)+\epsilon(x_2)|$ to find the smallest
absolute deviation is a good practical choice.  The precision of the
resulting approximation is
\begin{equation}
\frac{| \epsilon_N(x_1)-\epsilon(x_2)|}{|\epsilon_N(x_1)+\epsilon(x_2)|}
\end{equation}
over the desired range, so it is straightforward to pick a desired
accuracy for a known range simply by varying $N$. We illustrate an
example of a Zolotarev approximation in Fig.~\ref{fig:zol}.

Incidentally, the Zolotarev formula can be used to compute
$1/\sqrt{D^\dagger D}$, needed in the RHMC algorithm for simulating a single
flavor of Wilson-type fermions. The OpenQCD package does this
for the strange quark in its $2+1$ flavor simulations.

Evaluating the squared Hermitian overlap operator
$H^2=(\gamma_5 D)^2=D^\dagger D$
 underpins essentially all numerical work with the overlap.
Recall that $H^2$
commutes with $\gamma_5$ and  its eigenvectors can be chosen with
definite chirality. 
It is convenient to make use of the chiral projections ($P_\pm = \frac{1}{2}(1\pm \gamma_5)$)
so that the massive squared Hermitian overlap operator, with the usual
convention for the mass terms, is $H(m)^2= H_+(m)^2 + H_-(m)^2$ where
\begin{equation}
H^2_\pm(m) =  P_\pm H^2(m) P_\pm=2\left[ \left( \frac{r_0}{a}\right)^2
  -\frac{m^2}{4}\right] P_\pm (1\pm\epsilon(h))P_\pm+m^2 P_\pm \ .
\label{eq:hm2c}
\end{equation}
Calculations of the eigenmodes and eigenvalues of $H$ can be found by using the  $m=0$ version
of Eq.~\ref{eq:hm2c}. The full propagator will need $m\ne 0$, of course. In either case the
source has to have a definite chirality.

Eigenmodes and eigenvalues of $H$ have a literature of their own.
I have had good success with two methods to find eigenmodes and eigenvalues of $H$. One is an
implementation of the Ritz variational method
as described by the authors of Ref.~\cite{Kalkreuter:1995mm}
and the other is 
the ``primme'' package of Refs.~\cite{doi:10.1137/050631574,doi:10.1137/060661910}.

 Two examples of problems
addressed by eigenvalues of $H$ are the topological susceptibility, which can be computed
just by counting  zero modes of $H$, and calculations of parameters of the chiral Lagrangian
which describes the low energy limit of QCD. 

Overlap fermions are particularly suited to a limit of QCD called
the ``epsilon regime''   \cite{Gasser:1986vb,Gasser:1987ah}.
 This is a situation involving a finite volume $V=L^3$ and a quark mass
so light that $m_\pi L << 1$ while $ML>>1$ for all other hadrons. The partition function is dominated
by the pions. Everything else is massive and their contribution to the partition function is 
exponentially suppressed.
In that limit there is a connection between eigenvalues of the Dirac operator (typically in fixed
sectors of topology)
and random matrix theory. The first description of this connection was by
Shuryak and Verbaarschot \cite{Shuryak:1992pi}
and there is a large subsequent literature.
An old review by Damgaard \cite{Damgaard:2001ep}
is still a good introduction to the subject.
The most recent discussion of this topic is probably the 2019 FLAG 
review \cite{FlavourLatticeAveragingGroup:2019iem}.
In a sentence, the probability distribution of Dirac eigenvalues can be used to determine the condensate
and other low energy constants of the chiral Lagrangian.

 The advantage of using the overlap as
opposed to other formulations of lattice fermions is its  exact chiral symmetry,
which strongly constrains patterns in the spectrum -- zero modes are chiral, nonzero eigenvalue  modes are not.
This spares a lot of argument about how lattice artifacts affect results. (For example, is a small
eigenvalue eigenmode an approximation to a true zero mode, or not?)
 An issue with the program is that in the epsilon regime, finite size
corrections are power law  in $1/L$, rather than exponential
($\exp(-m_\pi L)$) as they are for conventional measurements. In the end, large volumes were needed to overcome
these large corrections.

Studies of hadronic observables in the ``p-regime'' ($m_\pi L \gg 1$) are done basically identically
to the ones using nonchiral actions. Many auxiliary calculations which one would do with a nonchiral fermion
(clover or Wilson), such as the determination of an additive shift in the fermion mass
or separate evaluations of $Z-$factors of quantities related by chiral rotations,
become unnecessary (except as checks that things have been coded correctly). The awkward struggles
between flavor and taste so
typical of staggered fermion calculations are simply absent, too.
I found that this simplified my analysis of data in the projects I did.

Of course, the Dirac propagator is the work horse of nearly all QCD calculations.
A flowchart of the construction of $\hat D(m)^{-1}\chi $ goes as follows:
Temporarily label $C=1-m^2/(4r_0/a)^2$, $S=m^2/C$. Then
\begin{enumerate}
\item Solve for $\psi= (H^2+S)^{-1} \chi$ using your favorite sparse matrix inversion routine.
\item 
Before subtraction, compute
\begin{equation}
D^{-1} \chi = D(m)^\dagger [H(m)]^{-2} \chi
    =\frac{1}{C} [(1-\frac{m}{2r_0/a})H \gamma_5 + m] \psi .
\end{equation}
\item 
Then the subtracted inverse Dirac operator (Eq.~\ref{eq:shift}) is
\begin{equation}
\hat D^{-1} \chi = \frac{1}{1-\frac{m}{2r_0/a}}[D^{-1} \chi - \frac{1}{2r_0/a}\chi].
\end{equation}
\end{enumerate}
It is useful to precondition the inversion using eigenmodes of $H$, if they are available.

Now for a few words about simulations with dynamical overlap fermions. 
All  the projects I know of used hybrid Monte Carlo (HMC) with the
$\Phi$  hybrid molecular dynamics (MD) algorithm.
The $\Phi$ algorithm
obtains the force on the gauge field from the derivative of the
pseudofermion effective potential $S_f = \Phi^* [M(U)^\dagger M(U)]^{-1}\Phi$
with respect to the vector potential
$A$ or $U$. 
Because the fermion matrix depends on the gauge field, the
pseudofermion term contributes the ``fermion force'',
\begin{equation}
  iF_{F,\mu n} = X^* \frac{\partial}{i\partial A_{\mu n}}[M(U)^\dagger M(U)] X.
\label{eq:fermforce}
\end{equation}
 As long as $M(U)$ changes smoothly with $U$
there is no problem evaluating this expression and one proceeds as one does in simulations
with some conventional fermion discretization.
 As we move along the molecular dynamics trajectory, however,
one of the eigenvalues of the kernel operator might change sign, signaling 
 the appearance or disappearance of a zero mode.
The topology of the underlying gauge configuration (as seen by the fermions) is trying to change.
When that occurs, the spectrum of the overlap operator $M(U)$ changes discontinuously.
This discontinuity generates an infinite fermion force.
Conventional MD stepping algorithms, such as
the leapfrog, fail, and  special treatment is called for.  In particular we
need a reversible MD treatment to carry us past the step.

A reversible molecular dynamics treatment of a discontinuous effective
potential was proposed by \cite{Fodor:2003bh}.  It is based on the
analogy with classical particle motion  in the presence of a step
barrier. Imagine a unit-mass particle moving in two spatial
dimensions $(x,y)$, and approaching a potential step, with $V(x,y)=0$, say, for
$x<0$ and $V(x,y)=V_0$ for $x>0$.
The component of momentum in the $y$ direction
is unchanged when the particle strikes the barrier, but there is an impulse
in the $x$ direction, which causes a discontinuous change in $p_x$. If the
barrier is too high, the particle ``reflects,'' $p_x \rightarrow -p_x$,
but if the barrier is low enough, the particle crosses the barrier (``refracts'') while $p_x$
changes. The change is given by energy conservation, $\frac{1}{2}p_x^2 = \frac{1}{2}(p_x')^2 +V_0$.

In the QCD simulation, we need to identify the Molecular Dynamics time at which one of the
eigenmodes of the kernel operator has a zero eigenvalue. We move along along the trajectory
monitoring the eigenvalues of $h$ until
we arrive at that ``crossing'' time. (This is not an extra 
expense, if eigenmodes of $h$ are being used to precondition the
evaluation of $\epsilon(h)$.) If we step ``across the boundary,'' so
 the minimum $\lambda$ changes sign, we have to go back
and adjust the integration step size until we land precisely on $\lambda=0$.
 Then we change the gauge field momentum $P$
discontinuously, reflecting or refracting according to the size of  the step in the potential $\Delta S_f$. 
After reflecting or refracting, we return to our usual leapfrog integration
until the end of the trajectory (or until we reach the next topological boundary).

Recall that the
analog of the energy was $\frac{1}{2}P^2 + S$.
In terms of the unit normal vector to the surface of discontinuity $N$, there are two possibilities for
changes in the gauge momenta, depending on the size of the momentum component
along the normal.

In compact notation we denote the inner product of two vector fields
in the Lie algebra of $SU(N)$ by
\begin{equation}
   N^* P = \sum_{n\mu} \mbox{Tr} (N_{n\mu}^\dagger P_{n\mu}) \ .
\end{equation}
Then, if $(N^* P)^2 > 2 \Delta S_f $, we have refraction:
\begin{equation}
\Delta P =-N \; (N^* P) + N \; {\rm sign} (N^* P) \;  \sqrt { (N^*P)^2
  -2 \Delta S_f}.
\end{equation}
Alternatively, if $(N^* P)^2 \le 2 \Delta S_f $, we have reflection:
\begin{equation}
\Delta P =-2  N  (N^* P) .
\end{equation}
 Here
 $\Delta S_f$ is the height of the discontinuity, which only arises from the fermionic part of the action.
It happens because one eigenvalue of the kernel action $h$ changes sign. (Recall Eq.~(\ref{eq:lowh}).)
Calling this eigenvector $\chi$, the (unnormalized)
 normal vector $N$ is  the matrix element
of the derivative of the kernel action with respect to $A$, computed in this special crossing state
\begin{equation}
 N(x,\mu)= \chi^* \frac{\partial h}{\partial A(x,\mu)} \chi .
\end{equation}

In practice, the exact step function of the overlap operator is replaced by
some approximation, such as a rational functional approximation.
 To obtain an accurate approximation to the step function, it continues to be necessary to
 project out low modes of the kernel operator, as in
Eq.~(\ref{eq:lowh}). This introduces another complication for HMC: we have to be able
to differentiate the projector.  Evaluating the
derivative requires determining the change in the projector under
small perturbations in $h$.  Labeling the projector on an eigenmode of $h(-r0)$
with eigenvalue $\lambda$ as $P_\lambda=
|\lambda \rangle\langle\lambda|$ , first order
perturbation theory gives \cite{Narayanan:2000qx}
\begin{equation}
\delta P_\lambda=\frac{1}{\lambda-h}(1-P_\lambda)\delta h P_\lambda+
P_\lambda \delta h\frac{1}{\lambda-h}(1-P_\lambda) .
\end{equation}
In the force, this term is combined with an analytic differentiation of the rational functional approximation,
using sources from which the low modes have been projected out, basically a mix of
 Eqs.~(\ref{eq:fermforce}) and (\ref{eq:lowh}).

All this business with reflection and refraction is morally equivalent to the problem of changing 
topology as seen in
simulations with actions which treat topology only approximately (meaning, everything 
else but overlap fermions).
What is different is that here, the barrier between topology changes is sharp and easy to understand.
As with other actions, simulations with overlap fermions have a hard time changing topology
and the problem gets worse as the lattice spacing decreases. It is very easy, though, to perform simulations
in sectors of fixed topology, simply by always reflecting from topological boundaries.
These data sets can used in the epsilon regime studies described above, to check predictions
for the properties of eigenvalues in sectors of fixed winding number.

The magical properties of overlap fermions allow simulations of an odd number of flavors without rooting.
For details, see Ref.~\cite{DeGrand:2006ws}.

Finally, let me make a few remarks about my own experiences with the overlap as a computational problem.
(This is described in more detail in Ref.~\cite{DeGrand:2000tf}.)
 I was never able to get anything useful out of an
overlap operator whose kernel $d$ or $h$ was a usual Wilson fermion with thin links.
It was too expensive to evaluate; $\epsilon_N$ needed very large $N$ and very small $c_{2l-1}$'s.
 The issue, I always thought,
was that while the spectrum of the overlap $D$ was a circle, that was not the case
 for the kernel Wilson action.
(An example was
 shown in Fig.~\ref{fig:circlew}.) I always felt that it was asking a lot of the computer to
map the crosses in Fig.~\ref{fig:circlew} onto the Ginsparg-Wilson circle. Actions with nearest 
neighbor and diagonal ($ \pm \hat i \pm  \hat j$) couplings produced
 approximately circular spectra to start with, and it was easier for the overlap operator to circularize them.
Smeared gauge connections plus a clover term further smoothed the action and allowed me to use
an $\epsilon_N$ with small $N$ and a more well-behaved range of eigenvalues of $h$.
I used eigenmodes of $h$ to precondition the calculation of $H$ and eigenmodes of $H$ to precondition
the calculation of $\hat  D^{-1}$.

Looking back, I am not sure that I found the most efficient algorithm.
 Every paper in
the literature using overlap fermions in a QCD context has its own long set of tricks to
tame the overlap and make it manageable.
Nothing ever produced an action which was less than about 50 times as expensive (in terms of the cost
to calculate $\psi = D \chi$) as the equivalent cost for a nonchiral action.

\section{Final thoughts}

At the end of this note I feel like I have mostly been writing about a
fascinating theoretical development which was a computational dead end.

A crude way to do history is to look up an article on the Inspire ({\tt https://inspirehep.net}) data base.
The articles posted there come with a histogram showing the number of citations per year.
One can mark the ``end of a subject'' by the time when the mean number of citations per year
falls below one, and the histogram breaks up
into separate blocks. For most of the overlap literature, that date is before 2015.

As far as I can tell from a literature search, only one lattice group doing large scale
simulations used overlap fermions (though there were several groups doing small scale simulations). That was JLQCD,
who started carrying out $N_f=2$ flavor simulations around 2006, with a $16^3\times 32$ volume
 and quark masses down to a nominal 20 MeV or $m_q=m_{strange}/6$ -- see Ref.~\cite{JLQCD:2006lvu}.
 A year later they were doing $N_f=2+1$ flavor
simulations. Their overlap program continued until about 2014, at which point they converted to
simulating Moebius domain wall fermions \cite{Noaki:2014sda}. These fermions do not have exact chiral symmetry, but
the additive mass renormalization they observed was about  1/2 MeV, which they argued was more
than adequate for their needs.

Advances in techniques also rendered many of the theoretical advantages of exact chiral symmetry
less important. For example, these days the topological susceptibility can be measured using
gradient flow \cite{Luscher:2010iy} from simulations with  nonchiral fermion actions, rather than by counting
zero modes. The condensate can be measured from the integrated spectral density of 
eigenvalues of a (nonchiral) Dirac operator \cite{Giusti:2008vb}.
And of course, most of the ``classic problems'' from the time of the development of the overlap
have been solved: measurements of the condensate or the pseudoscalar decay constant are
not so newsworthy any more. Groups have moved on to more complicated observables,
where many things have to be balanced against each other -- small lattice spacing, large volume, 
large statistics, multiparticle correlation functions, good chiral symmetry properties, and more.

I think that the organizers of this collection of articles wanted a small section about the overlap because it
represents an aspirational ideal: it is possible to encode a version of chiral symmetry
in a lattice fermion formulation which is exact even at nonzero lattice spacing.

And as we leave this review we can also leave QCD for a bigger question: is it possible to  construct
a lattice-regulated chiral gauge theory and thereby have a nonperturbative formulation of the Standard Model?
This has been an open problem for most of the lifetime of lattice QCD.
Much of the literature in this field starts with  modifications to domain wall or overlap fermions.
Perhaps the overlap story is not over quite just yet.

\begin{acknowledgments}
Most of what I learned about Ginsparg-Wilson fermions came via my collaborations with
Anna Hasenfratz,
Peter Hasenfratz,
Ferenc Niedermayer,
and
Stefan Schaefer.
I particularly want to say how much I learned from Peter Hasenfratz; I remember
how hard I worked to keep up with him. He was a great physicist and a good person.
\end{acknowledgments}

\bibliography{overlap}

\end{document}